\documentclass[apl,twocolumn,preprintnumbers,amsmath,amssymb,floatfix]{revtex4-2}
\usepackage{graphicx}% Include figure files
\usepackage{lmodern}
\usepackage{tabularx}
\usepackage{siunitx}
\usepackage{booktabs}
\usepackage{etoolbox}
\usepackage{epstopdf}
\usepackage{float}
\usepackage{placeins}
\usepackage{hyperref,graphicx}
\usepackage{dcolumn}% Align table columns on decimal point
\usepackage{bm}% bold math
\usepackage{color}
\usepackage{natbib}
\usepackage{amsmath}
\usepackage{amssymb}
\usepackage{multirow} % for row & col SPanning in tables
\usepackage{hyperref}
\usepackage{stackengine}
\usepackage{xcolor}
\usepackage{xr}
\hypersetup{colorlinks=true, linkcolor=blue, citecolor=red, urlcolor=magenta, pdftitle={DDPOs}, pdfauthor={Saurabh Ghosh}}

\usepackage[normalem]{ulem}

\begin{document}
%%%%%%%%%%%%%%%%%%%%%%%%%%%%%%%%
\title{Unveiling Insulating Ferro and Ferrimagnetism in Double-Double Perovskite Oxides}
%\title{First principles investigation on electronic, magnetic and optical properties of double-double perovskite oxides} 
\author{Monirul Shaikh$^{a}$}
\email{msk.phe@gmail.com}
\author{Fengyi Zhou$^b$}
\author{Sathiyamoorthy Buvaneswaran$^a$}
\author{Rajan Gowsalya$^a$}
\author{Trilochan Sahoo$^a$}
\author{Duo Wang$^b$}
\email{duo.wang@mpu.edu.mo}
\author{Saurabh Ghosh$^a$}
\email{saurabhghosh2802@gmail.com}
\affiliation{$^a$Department of Physics and Nanotechnology, SRM Institute of Science and Technology, Kattankulathur 603 203, Chennai, India}
\affiliation{$^b$Faculty of Applied Sciences, Macao Polytechnic University, Macao SAR 999078, China}
%%%%%%%%%%%%%%%%%%%%%%%%%%%%%
\begin{abstract}
The emergence of ferro- and ferrimagnetic behavior in insulating materials is uncommon, largely due to Hund's rules. Utilizing symmetry analysis, first-principles methods, and classical Monte Carlo simulations, \textcolor{black}{we report technologically important insulating ferro and ferrimagnetic double-double perovskite oxides. Our study predicts LaA$^{\prime}$MnNiO$_6$ (A$^{\prime}$ = V, Cr, Mn, Co, and Ni) as promising candidates for spintronic and optical applications exhibiting band gaps between 1.3 eV and 1.9 eV. We explain the mechanisms driving band gap openings and magnetic exchange interactions in these ferro and ferrimagnetic compounds. Monte Carlo simulations, together with state-of-the-art orbital-decomposed exchange parameter analysis, reveal intriguing variations in magnetic transition temperatures (up to 242 K) and the corresponding exchange mechanisms in all LaA$^{\prime}$MnNiO$_6$ compounds.}
In addition, we assess the thermodynamic and dynamic stability of these compounds to comment on the feasibility of these systems.
\end{abstract}
\maketitle
%%%%%%%%%%%%%%%%%%%%%%%%%%%%%

\section{INTRODUCTION}
In double perovskite oxides (DPOs) A$_2$BB$^{\prime}$O$_6$, the A-sublattices are occupied by alkaline-earth or rare-earth ions, while the B/B$^{\prime}$ sublattices are filled with transition metal (TM) ions, contributing to their functional properties \cite{king2010cation, das2008electronic}. Cation-ordered DPOs exhibit a wide range of atomic combinations at both A- and B-sites due to their structural and compositional flexibility~\cite{AALPatwod, king2010cation, buvaneswaran2023design}. The A- and B-sublattices can be arranged in different patterns, such as layered, columnar, or rock salt configurations \cite{shaikhfirst}. Recently, by utilizing the hybrid improper ferroelectric mechanism, we engineered a series of cation-ordered DPOs with promising multiferroic properties~\cite{CmemMat2021}. Through the application of machine learning techniques, we have identified the key factors that contribute to the stability of A-site layered, B-site rock salt DPOs~\cite{ghosh2022insights}.

While in double-double perovskite oxides (DDPOs), an emerging class of materials order 1:1 at both A- and B-sublattices, formulating to AA$^{\prime}$BB$^{\prime}$O$_6$ structures~\cite{solana2016double, solana2019, mcnally2017complex, li2018new}, 
three cation sites \textit{i.e.,} A$^{\prime}$, B- and B$^{\prime}$ are now occupied with TM ions. Moreover, in DDPOs, the A-sublattice orders in columns with different cation coordination than that of DPOs. These materials hold a very high degree of B-sublattice cation ordering in rock-salt and A-sublattice cation ordering in columns with a $a^+a^+c^-$ type octahedra rotations~\cite{mcnally2017complex}. These arrangements in TM ions dictate complex electronic structures with the potential of large magnetic interactions, thereby providing elevated magnetic transition temperatures to be operated for practical applications \cite{shaikh2024design}.

The first synthesized CaFeTi$_2$O$_6$ DDPO crystallizes in a tetragonal centrosymmetric space group \textit{P4$_2$/nmc} with a 10-fold Ca-coordination, a tetrahedral Fe-coordination, and another coplanar Fe-coordination~\cite{leinenweber1995high}. \textcolor{black}{In recent years, A-site ordered CaMnTi$_2$O$_6$ DDPO with similar coordination of CaFeTi$_2$O$_6$ at the A-sublattices draws attention for its ferroelectric distortion with a large bandgap ~\cite{aimi2014high, gou2017site, li2018new, antonov2020electronic}. However, CaMnTi$_2$O$_6$ DDPO shows an antiferromagnetic ground state with magnetic transition temperature $\sim$10K, limiting its applications for spintronics at an elevated temperature\cite{gou2017site}.}

La$_2$MnNiO$_6$ holds promise for an elevated transition temperature together with rich electronic structures \cite{das2008electronic}. In addition, the La$_2$MnNiO$_6$ compound is a ferromagnetic insulator. We, therefore, choose to work with La$_2$MnNiO$_6$ for designing ferro and ferrimagnetic DDPOs within the insulating phase with high transition temperatures. Ferro and ferrimagnetic spin ordering with semiconducting band structures are immensely important for a wide range of applications. The ferro/ferrimagnetic insulators are crucial for realizing dissipationless optoelectronic and spintronic devices~\cite{meng2018strain, wakabayashi2019ferromagnetism, nafradi2020tuning, yeats2017local}. \textcolor{black}{The ever-increasing demand for information storage and transport with reduced energy consumption of electronic devices is the key for microelectronic industries. However, ferromagnetic insulators (FMIs) are rare due to Hund's rule. Typically, neighboring species' electron spins align antiferromagnetically to minimize the system's energy. Hence, insulation and ferromagnetism are often found incompatible in a single set of materials \cite{zhao2023cacu3mn2te2o12, meng2018strain}. As a consequence, the FMIs are limited in nature \cite{nafradi2020tuning, meng2018strain, feng2014high}.}

\textcolor{black}{In this study, by employing symmetry operations, \textit{ab initio} density functional theory computations, and Monte Carlo simulations, we identify a set of ferro and ferrimagnetic insulating DDPOs. Together with A-site cation ordering, chemical substitution, and the $a^+a^+c^-$ tilt pattern \cite{shaikh2024design} in La$_2$MnNiO$_6$ compound provide us with LaA$^{\prime}$MnNiO$_6$ (A = V, Cr, Mn, Co, and Ni) DDPOs that exhibit forbidden energy gaps between 1.3 eV to 1.9 eV. These energy gaps are considered suitable for absorption of the solar spectrum \cite{buvaneswaran2023design}.  
The classical Monte Carlo simulations unveil that LaA$^{\prime}$MnNiO$_6$ compounds exhibit significantly large transition temperatures in comparison to that reported in ref. \cite{guo2019nonlinear}. Further analysis, based on calculated orbital-decomposed exchange parameters, reveals the intriguing magnetic exchange mechanisms that are directly dictated by the nominal charge state, the crystal field environment, and the resulting orbital occupations.}
In addition, we elucidate the origins of band gaps opening and the root cause of enhanced magnetic transition temperatures using exchange mechanisms in these compounds. Finally, we study different spin configurations and the formation of these DDPOs with respect to their decomposition into possible stable compositions and their dynamic stability.
%%%%%%%%%%%%%%%%%%%%%%%%%%%%%%%%%%%%%%%%%%%%%
\section{METHODOLOGY}
We begin our simulations using $\textit{ab-initio}$ density functional theory calculations~\cite{DFT} as implemented in Vienna $\textit{ab initio}$ simulation package (VASP)~\cite{vasp} to find magnetic ground states of the DDPOs (see TABLE T1 of the Supplementary Materials for energy values). We consider the generalized gradient approximation (GGA) augmented by the Hubbard-$U$ corrections (GGA+$U$) to describe the exchange-correlation effect~\cite{anisimov1997first}. To consider $d-d$ Coulomb interactions, we employ $U_E$ ~\cite{Dudarev1} (= $U-J_H$, where $J_H$ is Hund’s exchange parameter) parameters of 3.0 eV for V-$d$, 3.1 eV for Cr-$d$, 3.4 eV for Co-$d$, 3.9 eV for Mn-$d$, and 6.0 eV for Ni-$d$ electrons ~\cite{CmemMat2021, shaikh2024design}. The Kohn-Sham equations are solved using the projector augmented wave (PAW) method~\cite{paw}. The exchange-correlation part is estimated by the PBEsol functional~\cite{PBEsol}. We carry out a $\Gamma$-centered 4 $\times$ 4  $\times$ 4 $k$-point mesh obeying the crystal symmetry for Brillouin zone (BZ) sampling. A cutoff energy of 520 eV is set for all calculations. All geometry relaxations are carried out until changes in the total energy between relaxation steps are within 1 × 10$^{-6}$ eV, and atomic forces on each of the atoms are smaller than 0.01 eV/\text{\AA}.

We perform phonon calculations on the fully relaxed structures using the finite difference method as implemented in VASP \cite{finite1}. To draw the geometry of our systems, we make use of the visualization for electronic and structural analysis (VESTA) tool \cite{vesta1}. The symmetry operations are performed with the help of the ISODISTORT tool~\cite{campbell2006isodisplace}.
For optical properties, we adopt a full-potential augmented plane wave with modified Becke-Johnson (MBJ) exchange-correlation functional \cite{suppmaterial2025} as implemented in the WIEN2k package \cite{blaha1990full}.

To evaluate the inter-site magnetic exchange parameters of the DDPOs, the full-potential linear muffin-tin orbital (FP-LMTO) code RSPt~\cite{wills2010full} is used.
We utilize the optimized crystal structure and the calculated magnetic ground state configuration as the input. 
The magnetic force theorem~\cite{10.1016/0304-8853(87)90721-9} is used to extract exchange parameters: by introducing an infinitesimal spin deviation into the paired spin system and then mapping the corresponding energy variation to the sum of one-particle energy changes for the occupied states at the fixed ground state potential.
To ensure the convergence of the self-consistent calculation is sufficient to perform this state-of-the-art calculation, a $6\times 6\times 6$ grid of k-points is used for the BZ sampling.
% \textcolor{red}{The exchange-correlation functional and the on-site Hubbard-$U$ corrections are consistent with the \textcolor{purple}{\sout{other} ground state} calculations.}
% as GGA+U$_{\rm eff}$ with the U$_{\rm eff}$ are set to 3.1, 3.9, 3.0, 3.4, and 6.0 eV for Cr, Mn, V, Co, and Ni in LA$^\prime$MNO, respectively

The calculated exchange parameters are further used to construct an effective spin Hamiltonian, $$\mathcal{H}=-\sum_{i\neq j}J_{ij}e_i\cdot e_j,$$which is then used to perform classical Monte Carlo simulations using the UppASD code~\cite{skubic2008method}.
To obtain a precise value of magnetic transition temperature, the cumulant crossing method is employed~\cite{binder1981critical,binder1981finite}. Specifically, simulations with three lattice sizes, $13\times 13\times 13$, $15\times 15\times 15$, and $17\times 17\times 17$, are used with consideration of periodic boundary conditions.
An annealing process is simulated, starting at 500 K and gradually decreasing to 0 K, with a cooling rate that is denser around the respective phase transition temperatures. At each temperature, a 50,000-step simulation is performed to ensure that the system reached its equilibrium state.
In comparison, we also use  the calculated exchange parameters to evaluate the ordering temperature in the mean-field approximation (MFA)~\cite{kvashnin2015exchange}: $T_c^{\rm MFA}=\dfrac{2J_0}{3k_B}$, where $J_0=\sum_j J_{0j}$ corresponds to the sum of the exchange interaction energies.
%%%%%%%%%%%%%%%%%%%%%%%%%%%%%%%%%%%%%%%%%%%%%%
\section{RESULTS AND DISCUSSIONS}
\subsection{Structural and magnetic properties}
%%%%%%%%%%%%%%%%%%%%%%%%
%%%%%%%%%%%%%%%%%%%%%%%%%%%%%%%%%%%%%%%%%%%%%%
\begin{figure}
\centering
%\subfloat
\includegraphics[width=\linewidth]{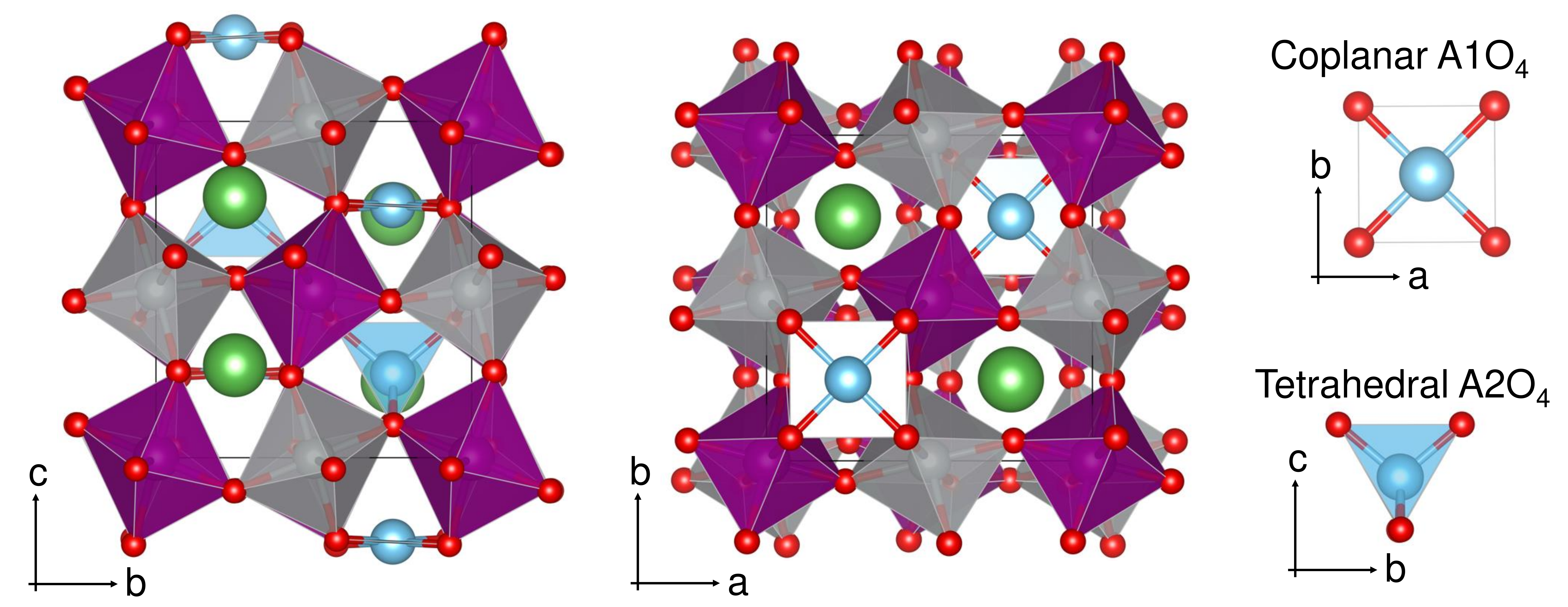}\vspace{-0pt}
\caption {(Color online) Crystal structure projections of LaA1$_{0.5}$A2$_{0.5}$MnNiO$_6$ along $a$-axis (left) and along $c$-axis (middle). A$^\prime$-sites are further split into A1O$_4$ coplanar and A2O$_4$ tetrahedral environments (right). The La-, A1/A2-, Mn-, Ni-, and O-atoms are described by green, blue, magenta, silver-color, and red spheres respectively.} 
\label{Structures}
\end{figure}
%%%%%%%%%%%%%%%%%%%%%%%%%%%%%%%%%%%%%%%%%

Figure \ref{Structures} illustrates the crystal structure of relatively new double-double perovskite oxides. In our DDPOs the A-, B-, and B$^{\prime}$-sites are kept fixed with La-atoms, Mn-atoms, and Ni-atoms respectively while A$^{\prime}$-site is varied with V-, Cr-, Mn-, Co-, and Ni-atoms. The A-site (La-atoms) shows 10-coordinated polyhedra while A$^{\prime}$-site is found to exhibit two different types of polyhedra. The first one shows a coplanar (CP) A1O$_4$ geometry and the second one displays a tetrahedral (TH) A2O$_4$ geometry. These inequivalent geometries order in a checkerboard type ordering within the A$^{\prime}$-site.
%%%%%%%%%%%%%%%%%%%%%%%%%%%%%%

The Mn- and Ni- at the B-sites form BO$_6$ octahedra almost equally tilted along the crystallographic $c$-axis as shown in Figure \ref{Structures}. The detailed crystal parameters are shown in TABLE T2 and TABLE T3 of the Supplementary Materials. It crystallizes in a tetragonal centrosymmetric $P4_2/n$ symmetry, which is a subgroup of $P4_2/nmc$ of the first synthesized CaFeTi$_2$O$_6$ DDPO \cite{leinenweber1995high}. Our observations are consistent with the recent experimental reports~\cite{solana2016double, solana2019, solana2021, mcnally2017complex, ji2023cafefenbo}.  
%%%%%%%%%%%%%%%%%%%%%%%%%%%%%

%======================================
Next, we investigate the stability of the working DDPOs from various spin configurations within the collinear picture. In LaA$^{\prime}$MnNiO$_6$ DDPOs, the A$^\prime$-site is adapted within the cavity of Mn/NiO$_6$ octahedra. Therefore, the nearest-neighbor distances are reduced effectually. As a result of that, the structure exhibits a complex magnetism within the collinear spin configuration. To single out the proper magnetic ground state for each compound, we exercise a systematic study by considering all possible collinear spins between A$^\prime$ site, the Mn-site, and the Ni-site as presented in Figure S1 of the Supplementary Materials. 

The magnetic ground states for LaCrMnNiO$_6$ (LCrMNO) and LaMn$_{A^\prime}$Mn$_B$NiO$_6$ (LMMNO) are found to be ferromagnetic. While LaVMnNiO$_6$ (LVMNO), LaCoMnNiO$_6$ (LCoMNO), and LaNi$_A$MnNi$_B{^\prime}$O$_6$ (LNMNO) show complex ferrimagnetic orderings. In the case of LVMNO and LNMNO the magnetic ground state is found to be the ferrimagnetic configuration of type-xiii in Figure S1. In this configuration, all A$^{\prime}$-sites (here A$^{\prime}$ = V and Ni) and Mn-sites move upward while Ni-sites are in opposite directions between two adjacent layers. In the case of LCoMNO the ground state magnetic ordering is identified to be type-xi in Figure S1. Here both Mn- and Ni-sites align ferromagnetically, while within A$^{\prime}$-site ( A$^{\prime}$ = Co) a G-type antiferromagnetic ordering is observed. The magnetic energies are calculated about the ferromagnetic spin configuration. Other collinear spin orderings are found to be stable within an energy window of $\sim$ 0.4 eV/f.u., as shown in TABLE T1 of the Supplementary Materials. The electronic structures corresponding to the magnetic ground states are investigated in the following section. 

\begin{table*}
\setlength{\tabcolsep}{10 pt} % horizontal spacing (column separation)
\renewcommand{\arraystretch}{1} %vertical spacing (row separation)
\caption{\label{tab:1} Magnetic moments, oxidation states, and band gaps of LaA$^{\prime}$MnNiO$_6$ DDPOs calculated using GGA+U+MBJ approach.}
\begin{tabular}{cc|ccc|ccc|c}
\hline
\hline
 & Systems & \multicolumn{3}{c|}{Magnetic moments ($\mu_B$) of}   &    \multicolumn{3}{c|}{Charge state of}  & Band gap \\
     &    &  A$^{\prime}$-site  & B-site & B$^{\prime}$-site & A$^{\prime}$-site  & B-site & B$^{\prime}$-site & E$_g$ (eV)\\
\hline
FMIs & LCrMNO & 2.67 & 2.88 & 1.85 & 3+ & 4+ & 2+ &2.01\\
     & LMMNO & 3.69 & 2.96 & 1.84 & 3+ & 4+ & 2+ &1.77\\
\hline
      & LVMNO & 0.91 & 3.66 & 1.87 & 4+ & 3+ & 2+ &1.92\\
FiMIs & LCoMNO & 2.08 (CP), 3.26 (TH) & 2.94 & 1.83  & 3+ & 4+ & 2+ &1.43\\
      & LNMNO & 1.37 (CP), 2.44 (TH) & 2.85 & 1.84  & 3+ & 4+ & 2+ &1.27\\
      \hline
      \hline
\end{tabular}
\end{table*}
%%%%%%%%%%%%%%%%%%%%%%%%%%%%%
%======================
\subsection{Electronic structures and optical transitions}
%======================

We perform the electronic structure calculations of our DDPOs through the density-functional theory within the GGA+U (see \textcolor{black}{Figure S2} of the Supplementary Materials for GGA+U electronic structures) and GGA+U+MBJ as implemented in VASP \cite{vasp} and WIEN2k package \cite{blaha1990full} respectively. The possible optical
 applications in these compounds are evaluated by calculating their respective
 absorption coefficients, dielectric coefficients, and are discussed in the Supplementary Materials. Our study identifies that the predicted compounds have a better possibility to be used in the UV-visible spectra as compared to the CaMnTi$_2$O$_6$. It is worth mentioning that using GGA+U+MBJ approach, we obtained almost the same energy band gap for the CaMnTi$_2$O$_6$ system \cite{antonov2020electronic} as shown in Figure S4 of the Supplementary Materials. Based on magnetic ground state and electronic structures, our compounds can be classified into two groups (1) ferromagnetic insulators (FMIs) and (2) ferrimagnetic insulators (FiMIs). The magnetic moments, oxidation states, and semiconducting band gap values of these DDPO compounds are listed in TABLE \ref{tab:1} 
\subsubsection{\textcolor{black}{Ferromagnetic insulators}}
\begin{figure*}
\centering
\includegraphics[width=\linewidth]{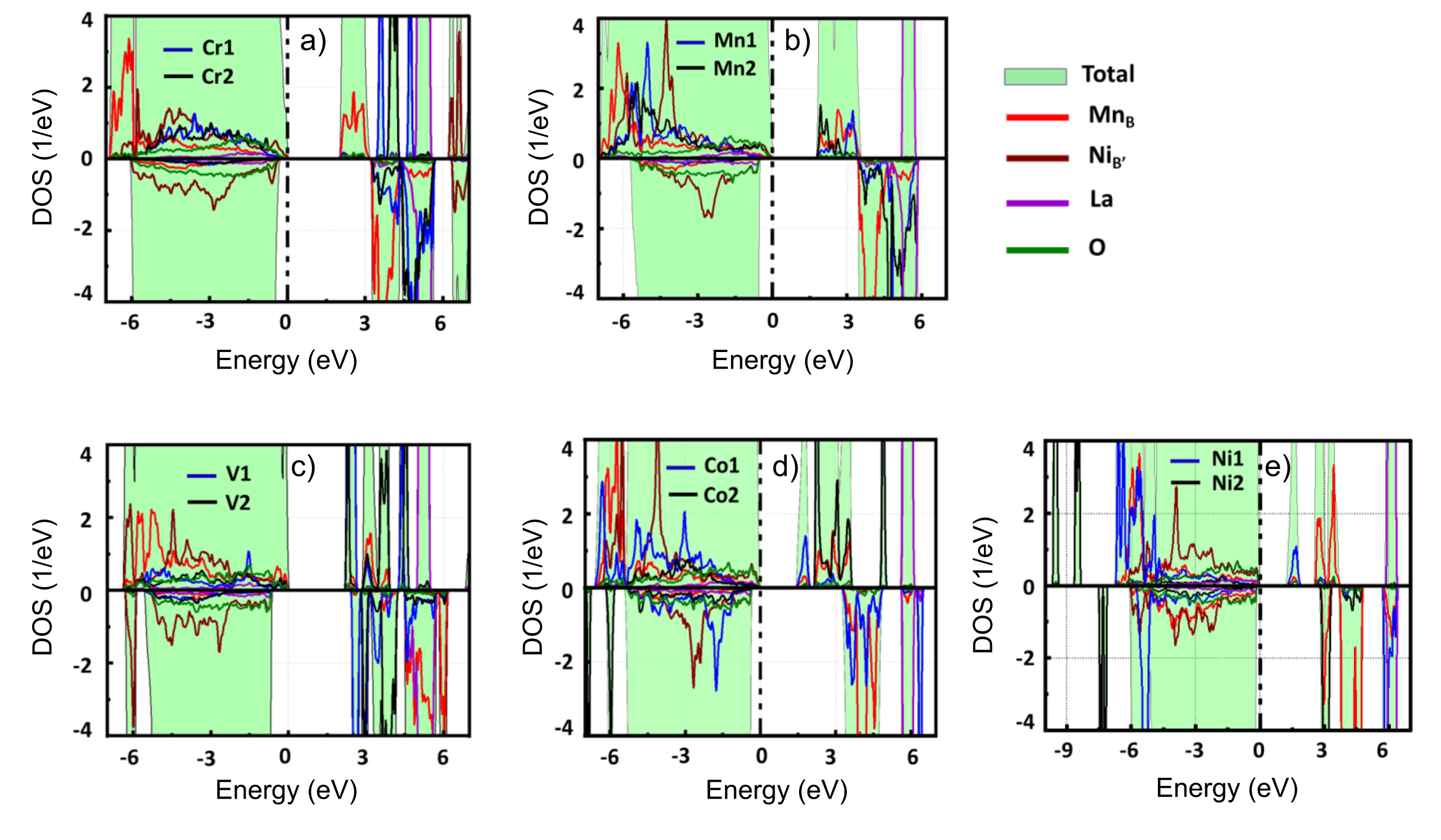}\vspace{-0pt}
\caption {Calculated DOS using the GGA+U+MBJ function for (a) LaCrMnNiO$_6$ (b) LaMnMnNiO$_6$ (c) LaVMnNiO$_6$ (d) LaCoMnNiO$_6$ and (e) LaNiMnNiO$_6$ respectively. Here shaded light green, red, black, purple, and dark green color lines represent the contribution of total, Mn$_B$, Ni$_{B'}$, La-, and O-atom respectively. The zero line in $x$-axis reads the valence band maxima.} 
\label{Figure3}
\end{figure*}

\textbf{LaCrMnNiO$_6$}: Figure \ref{Figure3}a shows the electronic structure of LaCrMnNiO$_6$ (LCrMNO). \textcolor{black}{Investigation of the partial DOS of LaCrMnNiO$_6$ reveals that both Mn-$d$ and Ni-$d$ orbitals split up into $t_{2g}$, and $e_g$ states due to octahedral surroundings in MnO$_6$, and NiO$_6$. The valance band up-spin channel (VBU) near the Fermi level ($E_F$) is primarily occupied by the hybridization of Ni$^{2+}$-$t^3_{2g\uparrow}e^2_{g\uparrow}$, Mn$^{4+}$-$t^3_{2g\uparrow}$ and Cr$^{3+}$-$e^2_{g\uparrow}t^1_{2g\uparrow}$ for TH geometry while $d_{xz}^1d_{yz}^1d_{z^2}^1$ for CP environment with mixing of O-$2p$ states. In comparison, the valance band down-spin channel (VBD) is only contributed by the Ni$^{2+}$-$t^3_{2g\downarrow}$ with strong mixing of O-2$p$ states. The Mn$^{4+}$-$e^0_{g\uparrow}$ state occupies the first energy band from the E$_F$ level in the conduction band up-spin channel (CBU). The conduction band down-spin channel (CBD) is predominantly occupied by the overlapping of Mn$^{4+}$-$t^0_{2g\downarrow}e^0_{g\downarrow}$ and Cr-states. Because the gap between the VB and CB up-spin channels is less than that of the down-spin channel, the initial optical transitions may happen from the filled mixing of Ni$^{2+}$-$t^3_{2g\uparrow}e^2_{g\uparrow}$, Mn$^{4+}$-$t^3_{2g\uparrow}$ and Cr-states to the empty Mn$^{4+}$-$e^0_{g\uparrow}$ state with O-2$p$ state mixing.} The band structure of this material in \textcolor{black}{Figure S3} in the Supplementary Materials reveals a semiconducting behavior with an indirect small band gap (2.01 eV) between high symmetry points Y and $\Gamma$.
%The filled Mn-$t_{2g}$ bands are located between -6.5 eV and Fermi level (E$_F$) in the up spin channel (USC). Mn-$t_{2g}$ in the down spin channel (DSC) is entirely empty. While Mn-$e_{g}$ in both the spin channels are found above the Fermi level. This suggests a nominal charge state of Mn$^{4+}$ ($t^3_{2g}e^0_g$) with a local moment of 2.88 $\mu_B$/Mn as shown in TABLE \ref{tab:2}. Ni-$t_{2g}$, and Ni-$e_g$ bands in the USC on the other hand lie between Mn-$t_{2g}$ and the Fermi energy and show a strong hybridization with Mn-$d$ and O-$p$.  While Ni-$t_{2g}$ bands in the DSC are localized between O-$p$ and the Fermi level. The $e_g$ levels in the DSC are located above E$_F$. This DOS along a local moment of 1.85 $\mu_B$/Ni indicates a nominal charge state of Ni$^{2+}$ ($t^6_{2g}e^2_g$). Further, the density of states (DOS) analysis of Cr-suggests that a local moment of 2.67 $\mu_B$/Cr with filled $d$-orbitals in the USC exhibits a nominal charge state of Cr$^{3+}$ ($d^3$) is in a high spin state and opens up a semiconducting gap (E$_{g}$) of 2.01 eV. Further, electronic band structure calculation reveals that the semiconducting gap is indirect as shown in Figure S1 of the Supplementary Materials.   

\textbf{LaMn$_{A^\prime}$Mn$_B$NiO$_6$}: Figure \ref{Figure3}b presents the electronic structure of LaMn$_{A^\prime}$Mn$_B$NiO$_6$ (LMMNO). The octahedrally coordinated Mn$_B$-$d$, and Ni-$d$ exhibit a band structure analogous to LCrMNO-system. However, the Mn$_{A^\prime}$-site shows a different electronic structure. \textcolor{black}{The CP Mn-states dictate an electronic configuration $d_{xz}^1d_{yz}^1d_{z^2}^1d_{xy}^1$ while TH Mn-states are found to exhibit an electronic configuration of $e_g^2t_{2g}^2$. The CBU and CBD channels are mainly occupied by overlapping of Mn$_{A^\prime}$ and Mn$_B$ with O-$2p$ state. It has a smaller band gap (1.77 eV) than A$^\prime$ = Cr material, permitting it to absorb more visible light.} 
%%%%%%%%%%%%%%%%%%%%%%%%%%%%%%%%%%%%%%%%%%%%% 

%%%%%%%%%%%%%%%%%%%%%%%%%%%%%%%%

\subsubsection{\textcolor{black}{Ferrimagnetic insulators}}
\textbf{LaVMnNiO$_6$}: The lowest magnetic configuration of LaVMnNiO$_6$ (LVMNO) is found to be ferrimagnetic ordering as described in Figure S1 (xiii) of the Supplementary Materials. The corresponding electronic structure is shown in Figure \ref{Figure3}c. In the LVMNO compound, in the case of V-atoms, we observe almost equal magnetic moments of 0.91 $\mu_B$/V for coplanar (V1-atom) and tetrahedral (V2-atom) geometries, respectively. The spin-resolved density of states shows that both the V-atoms contain a single unpaired electron. This clearly suggests that V-$d^1$ is in 4+ oxidation state. From the analysis of DOS, VBU is mainly contributed by the Mn$^{3+}$, V$^{4+}$, and Ni$^{2+}$ with strong mixing of O-$2p$ state. Contrast VBD is only contributed by the Ni$^{2+}$ with a strong mixing of O-2$p$ state. The CB up and down spin channels are mainly contributed by the hybridization of V$^{4+}$ and Mn$^{3+}$ with O-2$p$.
Mn-$t_{2g}$ bands below the Fermi level are found between -6.2 eV and E$_F$ in the up-spin channel. The band structure calculation uncovers a semiconducting gap (E$_{g}$) of 1.92 eV as shown in \textcolor{black}{Figure S3} in the Supplementary Materials for LVMNO. In this material, the optical transitions can mainly happen between the hybridization of Mn$^{3+}$, V$^{4+}$ and Ni$^{2+}$ to the hybridization of V$^{4+}$ and Mn$^{3+}$ with O-2$p$ state. \textcolor{black}{In a direct band gap semiconductor, the transition of electrons is permitted by both energy and momentum conservation. Hence, the transition process is efficient. This efficient optical transition is highly favorable for optoelectronic applications such as light-emitting diodes and photodetectors. Furthermore, the pronounced V–Mn orbital hybridization near the Fermi level enhances the joint density of states and can boost the probability of optical transitions, as shown in Figure \ref{Figure3}c. Our electronic structure and absorption coefficient analysis of LVMNO shows a strong absorption in the visible region (Figure S4 of the Supplementary Materials). These features collectively underscore the potential of this DDPO material for advanced optoelectronic and solar energy harvesting applications.}
\par
\textbf{LaCoMnNiO$_6$}: The ground state ferrimagnetic DOS of LaCoMnNiO$_6$ (LCoMNO) is shown in Figure \ref{Figure3}d. In the magnetic ground state of LCoMNO the coplanar Co-atoms interact antiferromagnetically with the tetrahedral Co-atoms. While ferromagnetic interactions are found between Mn- and Ni-sites. In the LCoMNO compound, in the case of Co-atoms, we find two different magnetic moments i.e., 2.08 $\mu_B$/Co and 2.94 $\mu_B$/Co for coplanar (Co1-atom) and tetrahedral (Co2-atom) geometries, respectively. The partial density of states analysis shows that Co-atoms hybridize strongly with the O-2$p$. \textcolor{black}{The electronic structure and the magnetic moments of CP Co-atom indicate that it obeys an electronic configuration of $d_{xz}^2d_{yz}^2d_{z^2}^1d_{xy}^1$ whereas for TH geometry it follows $e_g^3t_{2g}^3$ electronic configuration. The disparity between the two different magnetic moments is attributed to two different geometries and hence the electronic structures. This implies a 3+ oxidation state for the Co-atoms. The crystal field splitting and the partial density of states of Mn-$d$ and Ni-$d$ in the case of LCoMNO compound uncover electronic properties similar to LCrMNO compound. The electronic band structure calculation for LaMn$_{A^\prime}$Mn$_B$NiO$_6$ exhibits an indirect semiconducting gap of 1.43 eV as shown in \textcolor{black}{Figure S3} in the Supplementary Materials. In the CBU the first energy band near the E$_F$ level is filled by the Co1 and Mn$^{4+}$, whereas Co2 and Mn$^{4+}$ occupy the second. Both bands are strongly correlated with the O-2$p$ state. In this material, the optical transitions mainly can happen via the overlapping of Co, Ni$^{2+}$ and Mn$^{4+}$ with O-2$p$ states to the hybridization of Co and  Mn$^{4+}$ states.} 

\textbf{LaNi$_A$MnNi$_{B^\prime}$O$_6$}:
The ground state ferrimagnetic DOS of LaNi$_A$MnNi$_{B^\prime}$O$_6$(LNMNO) is shown in Figure \ref{Figure3}e. In the LNMNO compound, in the case of Ni$_A$-atoms, we notice two different magnetic moments i.e., 1.37 $\mu_B$/Ni$_A$ and 2.44 $\mu_B$/Ni$_A$ for coplanar (Ni1-atom) and tetrahedral (Ni2-atom) environment, respectively. \textcolor{black}{The partial density of states analysis shows that Ni-atoms hybridize strongly with the O-2$p$. In the case of TH geometry, it gives a fully filled $e_g^4$ state and a partially filled $t_{2g}^3$ state. In case of CP environment, it offers us a completely filled $d_{xz}^2d_{yz}^2$ and half-filled $d_{z^2}^1d_{xy}^1$ whereas the extra electron goes to the O-2$p$ and forms a ligand-hole recombination~\cite{gowsalya2023designing}. This, in turn, leaves an oxidation state 3+ for the Ni$_A$-atoms. The crystal field splitting and the partial density of states of Mn-$d$ and Ni-$d$ in the case of the LNMNO compound are the same as LCrMNO compound. The band structure plot in \textcolor{black}{Figure S3} in the Supplementary Materials reveals that LNMNO is an indirect band gap semiconductor of E$_g = 1.27$ eV.} The CBU first energy band close to the E$_F$ level is filled by an overlap of Mn and Ni1 mixed with O-$2p$. Mn occupies the subsequent band with a minor contribution from the O-2$p$ state. While the Mn and Ni2 states mainly contribute to the CBD.
   
%%%%%%%%%%%%%%%%%%%%%%%%%%%%%%%%%%%%%%%%
\subsection{Magnetic transition temperatures and exchange mechanism}
%%%%%%%%%%%%%%%%%%%%%%%%%%%%%%%%%%%%%%%%
%%%%%%%%%%%%%%%%%%%%
\begin{figure*}
    \centering
    \includegraphics[width=0.75\linewidth]{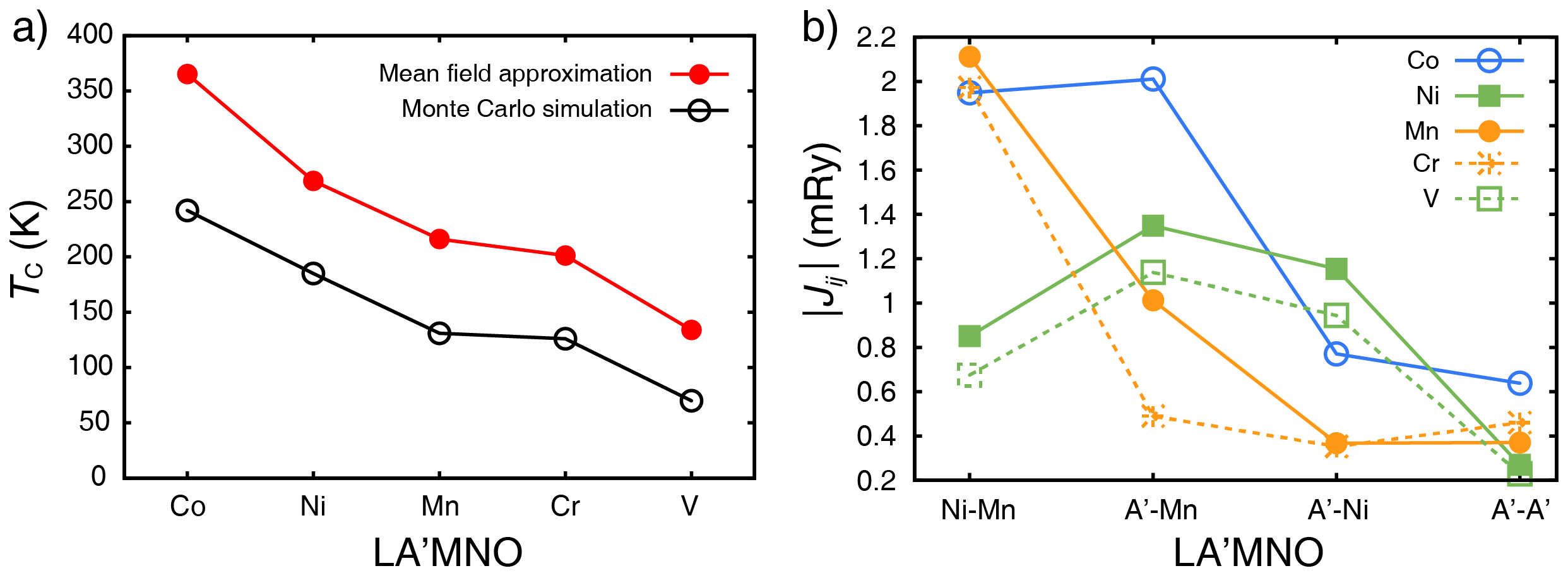}
    \caption{(a) \textcolor{black}{Calculated transition temperatures obtained from the mean field approximation and Monte Carlo simulation of the LA$^\prime$MNO (A$^\prime$ = Co, Ni, Mn, Cr, V).} (b) Averaged exchange parameters $J_{ij}$ for four important TM-TM pairs. Orange, blue, and green lines indicate the magnetic ground states as FM, A$^\prime$-site AFM, and B$^\prime$(Ni)-site AFM; dashed and solid lines further distinguish different compounds.}
    \label{fig:fig4}
\end{figure*}
%%%%%%%%%%%%%%%%%%%%%
By performing classical Monte Carlo simulation from high temperature to low temperature, and by adopting mean-field approximation (MFA), two sets of magnetic phase transition temperatures for the LA$^\prime$MNO compounds are obtained, as shown in Figure~\ref{fig:fig4}a.
With the A$^\prime$-site substitution, both data sets exhibit a wide range of $T_C$ variation, from 365 K to 134 K for MFA, and 242 K to 70 K for classical Monte Carlo results. 
It is expected that the simplified model, MFA, tends to overestimate transition temperatures, while the classical Monte Carlo simulations, by numerically mimicking the gradual cooling process, provide more accurate results.
Notably, the overall transition temperature trend obtained from the two methods is identical: the Co-substituted compound exhibits the highest, the Ni-based one the second, followed by the Mn-, Cr-, and V-based ones consecutively.
This interesting temperature variation introduced by a single-element substitution directly reflects the changes caused on the inter-site exchange couplings, and can therefore be well understood by analyzing the specific exchange mechanisms.

\begin{figure*}
    \centering
    \includegraphics[width=\linewidth]{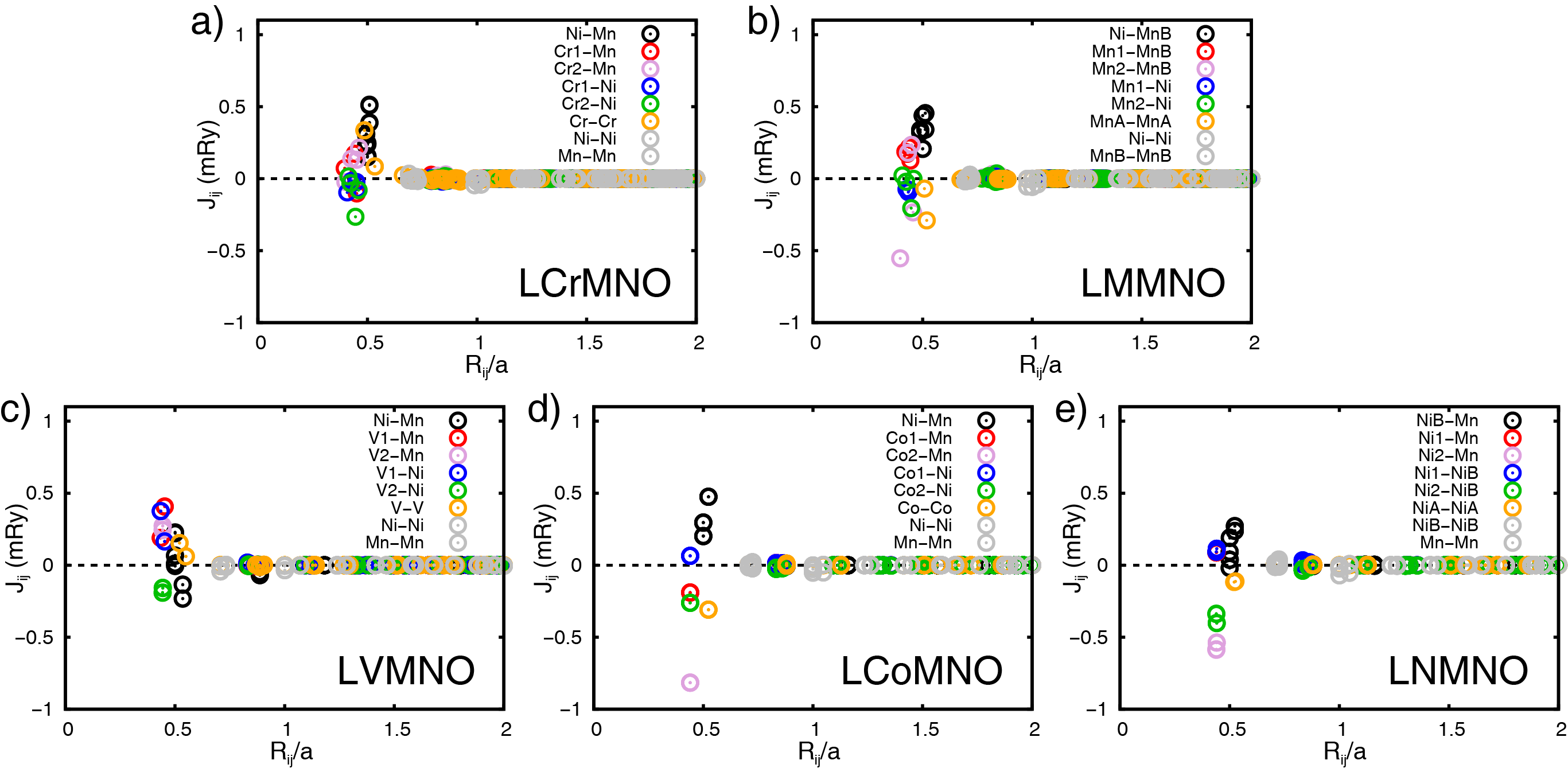}
    \caption{Calculated exchange parameters as a function of distance for all TM-TM pairs in (a) LCrMNO, (b) LMMNO, (c) LVMNO, (d) LCoMNO, and (e) LNMNO. Positive and negative values indicate FM and AFM coupling. A$^\prime$1 and A$^\prime$2 represent the substituted atoms located in coplanar and tetrahedral geometries.}
    \label{fig:fig5}
\end{figure*}

Averaged exchange parameters $J_{ij}$ for four important TM-TM pairs are shown in Figure~\ref{fig:fig4}b. It is easy to observe that, for any TM-TM pairs, the coupling strength decreases rapidly as the distance increases, indicating that only the first nearest neighbor couplings are important in the systems. Among them, four essential intersite coupling types were identified: Mn-Ni, A$^\prime$-Mn, A$^\prime$-Ni, and A$^\prime$-A$^\prime$.
To extract hidden features in these four couplings and to gain insight into the reasons for the transition temperature variation, we post-processed the data as follows: for each coupling type, we took the absolute value of its nearest and next-nearest neighboring $J_{ij}$ values, and then summed them up.
   
Interestingly, it is observed that the coupling patterns of these five compounds are directly connected to their respective magnetic structures. 
Specifically, with the ground state as FM, LMMNO, and LCrMNO exhibit a clear pattern where Ni-Mn coupling 
is substantially strong, while the other three couplings are nearly negligible.
In LCoMNO, which has the A$^\prime$-site AFM as the ground state, both Ni-Mn and A$^\prime$(Co)-Mn couplings are significant.
In contrast, in LNMNO and LVMNO, the B$^\prime$(Ni)-site AFM state induces two magnetic strongly coupled interactions: A$^\prime$-Mn and A$^\prime$-Ni. Calculated exchange parameters ($J_{ij}$) for the five LA$^\prime$MNO compounds are shown in Figure~\ref{fig:fig5}. The exchange mechanism is the essential physical quantity that not only determines the ground state magnetic structure but also explains the variation in these coupling strengths, thereby dictating the critical temperatures in magnetic systems.
Orbital-decomposed exchange parameters, along with a schematic illustration of orbital-level hopping, have proven to be effective tools for unraveling complex exchange mechanisms in various magnetic systems~\cite{PhysRevMaterials.5.054405,wang2022ab,PhysRevB.104.245410,zhou2024emergence}. In the following discussion, we will thoroughly analyze the exchange mechanism by combining these two approaches.

Foremost, the coupling shared by all compounds --- the B$^\prime$(Ni)-B(Mn) interaction --- is the first one that needs to be understood.
As shown in Figure~\ref{fig:fig4}b, when the moments on the two sites are aligned in a parallel manner, exhibited in LMMNO, LCrMNO, and LCoMNO, it results in a strong coupling strength of nearly 2.0 mRy. 
    
In contrast, if there are moments ordered anti-parallel, as in the B$^\prime$(Ni)-site AFM compounds LNMNO and LVMNO, the coupling strength drastically drops to around 0.7 mRy. This phenomenon can be well understood by analyzing orbital-decomposed interactions, as shown in \textcolor{black}{Figure S6}.
As discussed in section B, the nominal charge state of B$^\prime$-site Ni is 2+, the three-fold degenerate lower energy $t_{2g}$ level is fully occupied ($t_{2g}^6$), and the two-fold degenerate higher energy $e_g$ level is half-filled ($e_g^2$), with a total of eight $d$-electrons. 

Similarly, with a total number of three $d$-electrons, the Mn$^{4+}$ charge state has a half-filled $t_{2g}$ level and an empty $e_g$ level ($t_{2g}^3$$e_g^0$). A schematic picture of these occupation states and a respective hopping mechanism are presented in Figure~\ref{fig:fig8}a. When the spin system forms an FM alignment (upper panel), the $e_g$ electron at the Ni site can freely hop to the empty $e_g$ state at the Mn site, whereas hopping of the $e_g$-$t_{2g}$ form is not allowed due to the Pauli's exclusion principle. 
On the other hand, in an AFM alignment (lower panel), the majority spin ($\uparrow$) in the unpaired $e_g$ level, rather than the paired spin in the lower $t_{2g}$ level, is more energetically favorable to hop to the Mn $t_{2g}$ level due to the on-site Coulomb interaction. As a result, the two processes lead to clear exchange forms of FM $e_g$-$e_g$ and AFM $e_g$-$t_{2g}$.
Eventually, the overall exchange coupling is determined by the competition between the two. When the FM $e_g$-$e_g$ coupling is stronger than the AFM $e_g$-$t_{2g}$ coupling, as illustrated by the orbital-resolved $J_{ij}$ data in \textcolor{black}{Figure S6} in the Supplementary Materials for LCrMNO, LMMNO, and LCoMNO, the system exhibits FM Ni-Mn interaction. In cases where the $e_g$-$e_g$ coupling strength is diminished, such as LNMNO and LVMNO, the result is a nearly zero or even weak AFM $J_{total}$. Particularly, the weak AFM in the Mn$^{3+}$-based LVMNO is caused by the additional $e_g$ electron at the Mn site ($t_{2g}^3e_g^1$), which adds additional $d-d$ Coulomb repulsion on the $e_g$-$e_g$ hopping. It restricts the corresponding kinetic energy and therefore allows the FM coupling strength to be overcome by the AFM $e_g$-$t_{2g}$ coupling.

 \begin{figure*}
    \centering
    \includegraphics[width=0.8\linewidth]{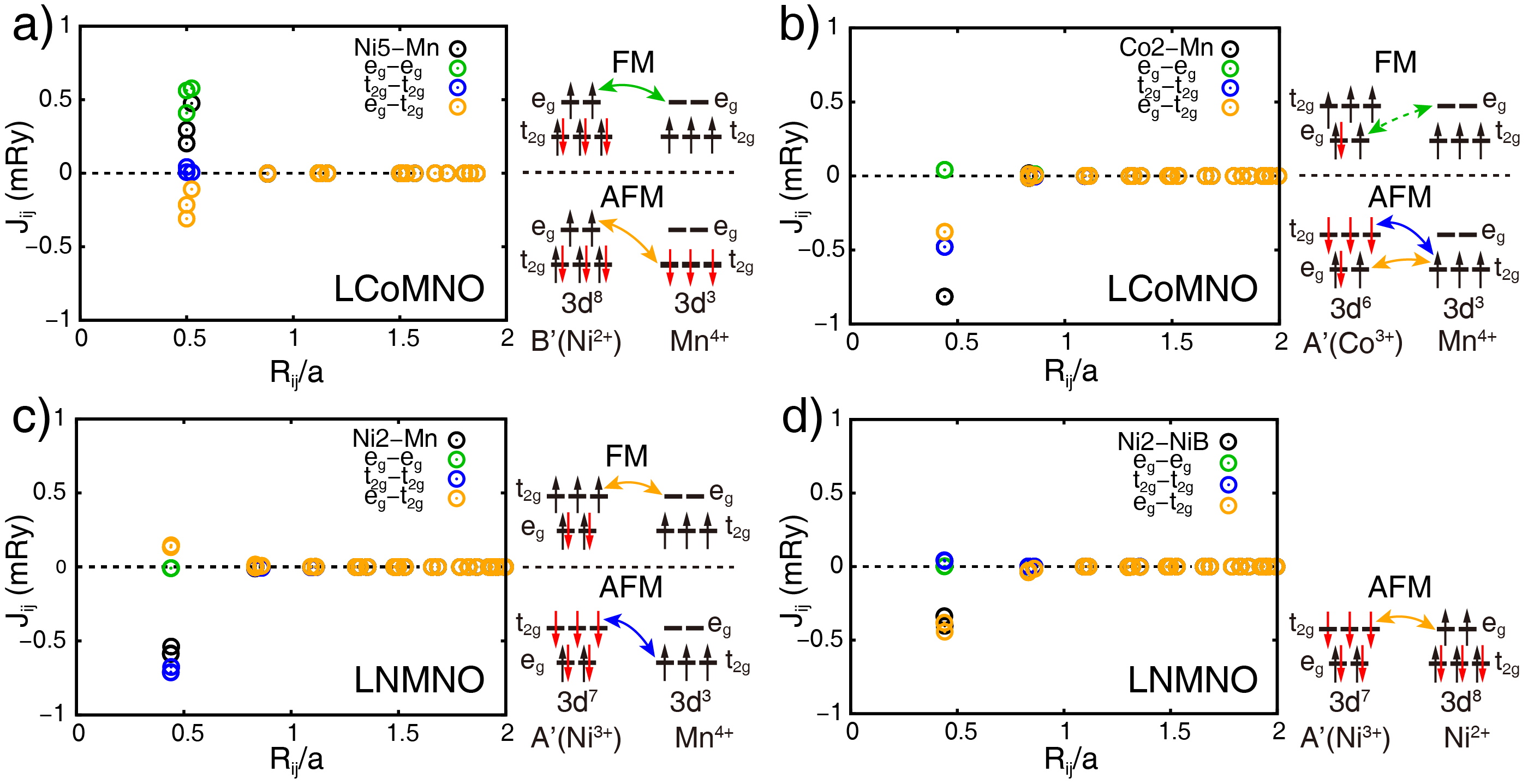}
	\caption{\label{fig:fig8} 
   \textcolor{black}{Calculated orbital-decomposed exchange parameters as a function of distance (left panel) and the corresponding schematic illustration of the orbital-dependent electron hopping mechanisms (right panel) for (a) B(Mn)-B$^\prime$(Ni), (b) B(Mn)-A$^\prime$(Co), (c) B(Mn)-A$^\prime$(Ni), and (d) B(Ni)-A$^\prime$(Ni) interactions in the representative compounds.}
   % (a) B$^\prime$(Ni)-B(Mn), (b) A$^\prime$(Co)-B(Mn), (c) A$^\prime$(Ni)-B(Mn), and (d) A$^\prime$(Ni)-B(Ni) interactions in the representative compounds.
   % The detailed mechanism for LCoMNO compound with the octahedral atoms pair as (a) B$^\prime$(Ni)-B(Mn) interaction and (b) the A$^\prime$(Co)-B(Mn) interaction connected to the tetrahedral A$^\prime$-site atoms, for LNMNO with the (c) A$^\prime$(Ni)-B(Mn) and (d) A$^\prime$(Ni)-B(Ni) interaction.
   }
    \label{fig:fig8}
    \end{figure*}

The coupling between the A$^\prime$ atom and others is relatively complex, as the A$^\prime$ atom varies in different compounds, and as the local environment around the A$^\prime$ atom in each compound includes both coplanar (A1) and tetragonal (A2) crystal fields. 
By comparing the coupling strengths on these two different atoms, we found that those on A2 
is significantly stronger. This is illustrated in Figure~\ref{fig:fig5}(a, b, d and e) with pink and green data, indicating a more prominent effect on critical temperatures. Therefore, in the following discussions, we will mainly focus on the couplings on the A2 atoms. The orbital-decomposed Heisenberg exchange of the tetrahedral A$^\prime$ atoms and B-site Mn atoms in LCrMNO and LMMNO are shown in \textcolor{black}{Figure S7} in the Supplementary Materials. 

As it is shown in Figure~\ref{fig:fig4}b, among A$^\prime$-Mn, A$^\prime$-Ni, and A$^\prime$-A$^\prime$ couplings, A$^\prime$-Mn is responsible for LCoMNO having the highest phase transition temperature in the five compounds.
With a nominal charge of 3+, the six $d$-electrons on the Co site are arranged in response to the local crystal field environment. In the coplanar environment, the electrons fill the orbitals with fully occupied $d_{xz}$, $d_{yz}$, and half-filled $d_{xy}$ and $d_{z^2}$ ($d_{xz}^2 d_{yz}^2 d_{xy}^1 d_{z^2}^1$), resulting in a magnetic moment of 2.0 $\mu_B$. While in the tetragonal environment, especially for those with a relatively small crystal field splitting energy, electrons can easily overcome the energy gap between the lower $e_g$ and higher $t_{2g}$ level, thereby forming a high-spin $e_g^3 t_{2g}^3$ state. A schematic picture and the calculated orbital-resolved $J_{ij}$ values are shown in Figure~\ref{fig:fig8}.
With an AFM A$^\prime$(Co)-Mn coupling, the $t_{2g}$ electron on the Mn site can easily hop to either $e_g$ or $t_{2g}$ level on the Co site, resulting in a negative $e_g$-$t_{2g}$ (yellow) and $t_{2g}$-$t_{2g}$ (blue) contribution. 
Since there is already a spin-down electron occupying the Co $e_g$ level, the $e_g$-$t_{2g}$ hopping is therefore restricted, resulting in a relatively smaller coupling magnitude (-0.38 mRy) compared to the $t_{2g}$-$t_{2g}$ one (-0.48 mRy). Together with a negligible $e_g$-$e_g$ contribution (0.04 mRy), the compound exhibits a significant AFM coupling of -0.82 mRy for a single A$^\prime$(Co)-Mn pair (black data in Figure~\ref{fig:fig8}b).
As a result, the summed-up Ni-Mn and A$^\prime$(Co)-Mn coupling strength reaches 1.95 mRy and 2.01 mRy, therefore giving LCoMNO the highest phase transition temperature of 242 K. The orbital-decomposed Heisenberg exchange of the tetrahedral A$^\prime$ atoms and B-site Ni atoms in all LA$^\prime$MNO are shown in \textcolor{black}{Figure S8} in the Supplementary Materials.
  
On the other hand, in LNMNO and LVMNO, the A$^\prime$-Mn and A$^\prime$-Ni coupling strengths are the critical factors that determine their $T_C$.
In the Ni-substituted compound, the nominal state of Ni$^{3+}$ in the tetragonal environment gives it a fully occupied $e_g$ state and a half-filled $t_{2g}$ state ($e_g^4 t_{2g}^3$).
    For a single A$^\prime$(Ni)-Mn coupling, when the spin alignment between A$^\prime$(Ni) and Mn becomes AFM, the majority spin ($\uparrow$) at the Mn $t_{2g}$-state can easily hop to the majority spin ($\downarrow$) at the Ni $t_{2g}$-state. Although there is a small FM compensation originating from the Mn($e_g^0$)-Ni($t_{2g}^{3,\uparrow}$) hopping, the overall A$^\prime$(Ni)-Mn coupling is a strong AFM with an average value of -0.56 mRy, as illustrated in orbital-decomposed data in Figure~\ref{fig:fig8}c.
        For the A$^\prime$(Ni)-Ni coupling, the B$^\prime$-site Ni atom consists of eight $d$-electrons in an octahedral environment, a fully occupied $t_{2g}$ level and a half-filled $e_g$ level is established ($t_{2g}^6 e_g^2$). The energy-favorable AFM hopping happened between the majority spins from the two half-filled states: the $t_{2g}^\downarrow$ at the A$^\prime$-site Ni and the $e_g^\uparrow$ at the B-site Ni, resulting in a total exchange parameter of -0.37 mRy (Figure~\ref{fig:fig8}d).
        As illustrated in Figure~\ref{fig:fig4}, these two interactions give LNMNO a summed-up coupling strength of 1.35 mRy and 1.15 mRy for A$^\prime$-Mn and A$^\prime$-Ni coupling, respectively. This successfully compensates for its coupling strength drop in the Ni-Mn interaction and gives the system the second-highest $T_C$ of 185 K.
    In LVMNO, the situation is similar; however, fewer electrons ($e_g^1$) occupy the A$^\prime$(V)-site, resulting in a significantly reduced coupling strength. The corresponding process is illustrated in \textcolor{black}{Figure S9} in the Supplementary Materials, exemplifying the lowest $T_C$ of 70 K.
    In comparison, due to the strong coupling of Ni-Mn and a noticeable variation in the A$^\prime$-Mn coupling, LMMNO and LCrMNO have $T_C$ values of 131 and 126 K, respectively.
%%%%%%%%%%%%%%%%%%%%%%%%%%%%%%%%%%%%%%
\subsection{Thermodynamic and dynamic stability}
The thermodynamic stability is one of the basic requirements for synthesizing a compound for practical applications. We, therefore, calculate the formation energies using plane wave pseudo-potentials for these systems by considering the decomposition reaction of DDPOs via the most probable reaction pathways by utilizing a linear programming problem clubbed with the grand canonical method~\cite{Akbarzadeh2007, CmemMat2021, DuoPRM}. The detailed methodology can be found in these references \cite{CmemMat2021, DuoPRM, shaikh2024design}, and the relevant equations and constituents (TABLE T5) are provided in the Supplementary Materials.   
%%%%%%%%%%%%%%%%%%%%%%%%%%%%%%%%%%%%%%%%%%%
\begin{table}
\caption{\label{tab:4} Formation energies of the double-double perovskite systems.}
\begin{tabular}{cc}
\hline
\hline
Double-double perovskite oxides & Formation energies (eV/f.u.)\\
\hline
CaMnTi$_2$O$_6$     & 1.15 \\
LaVMnNiO$_6$     & 0.87 \\
LaCrMnNiO$_6$    & 0.91\\
LaMn$_{A^\prime}$Mn$_B$NiO$_6$ & 1.49 \\
LaCoMnNiO$_6$    & 1.04 \\
LaNi$_{A^\prime}$MnNiO$_6$  & 1.10 \\ 
\hline
\hline
\end{tabular}
\end{table}
%%%%%%%%%%%%%%%%%%%%%%%%
Implementing the above methodology, the formation energies of the double-double perovskite oxides are shown in TABLE \ref{tab:4}. The formation energies emerged to be positive and are in the same order of magnitude as reported in a recent study on quadruple perovskite oxides \cite{DuoPRM}. It is to be noted that the formation energy for CaMnTi$_2$O$_6$ is found to be positive, which is again consistent with the synthesizing techniques \cite{aimi2014high}.

The dynamic stability of all the La${A^\prime}$MnNiO$_6$ compounds are examined using finite-difference method phonon calculations~\cite{finite1} as implemented in VASP code. The lattice dynamics of these compounds are investigated using their respective phonon dispersion plots as shown in \textcolor{black}{Figure S10} in the Supplementary Materials. Our phonon dispersion plots show three tiny imaginary frequencies near the $\Gamma$-point. The range of these frequencies lies between 0 to 2 $cm^{-1}$. Close analysis reveals that these tiny phonon dispersions are all acoustic and arise due to numerical errors in the calculations. Hence, La${A^\prime}$MnNiO$_6$ compounds are considered to be kinetically stable \cite{zhang2012two, mahata2017free}. 
%%%%%%%%%%%%%%%%%%%%%%%%
\section{Conclusion}
To conclude, we conduct first-principles DFT calculations, symmetry analysis, and Monte Carlo simulations to explore the ferro/ferrimagnetic insulating behavior within double-double perovskite oxides (DDPOs). Our findings can be summarized as follows:

\begin{itemize}
    \item \textbf{Electronic Structure and Band Gap Formation}:
    \begin{itemize}
        \item \textcolor{black}{La${A^\prime}$MnNiO$_6$ compounds (${A^\prime}$ = Cr, Mn, V, Co, and Ni) exhibit semiconducting band gaps between 1.3 eV and 2.0 eV, which should be ideal for optoelectronic applications.}
        \item \textcolor{black}{The band gap openings are explained through a detailed analysis of the electronic structures, where $d$-bands are positioned near the Fermi level, enhancing the possibility for visible light absorption.}
    \end{itemize}
    
    \item \textbf{Spintronic Applications}:
    \begin{itemize}
        \item \textcolor{black}{The semiconducting energy gaps, along with suppressed electron transfer in these compounds, make La${A^\prime}$MnNiO$_6$ DDPOs promising candidates for spintronic devices.}
    \end{itemize}
    
    \item \textbf{Magnetic Transition Temperatures and Exchange Mechanisms}:
    \begin{itemize}
        \item \textcolor{black}{Monte Carlo simulations reveal significantly high magnetic transition temperatures in La${A^\prime}$MnNiO$_6$ compounds (ranging from 365 K to 134 K), crucial for practical device performance at elevated temperatures.}
        \item The exchange mechanisms responsible for enhanced magnetic transition temperatures are elucidated through orbital-decomposed exchange parameters.
    \end{itemize}
    
    \item \textbf{Thermodynamic and Dynamic Stability}:
    \begin{itemize}
        \item Thermodynamic analysis reveals that these DDPOs can form under high temperature and pressure, confirming their feasibility for experimental synthesis.
        \item Dynamic stability calculations confirm the kinetic stability of the compounds, ensuring their persistence in real-world conditions.
    \end{itemize}

\end{itemize}
%Our theoretical study identifies five La${A^\prime}$MnNiO$_6$ compounds as critical for developing ferro/ferrimagnetic insulators. These compounds show remarkably high transition temperatures, semiconducting properties, and strong potential for optoelectronic and spintronic applications.
\par
%To conclude, we carry out first-principles DFT calculations, symmetry analysis and Monte Carlo simulations to discover the ferro/ferrimagntic insulators within the double-double perovskite oxides. We explain the band gap openings and optical transitions in these compounds by analysing the electronic structures. We unravel that  La${A^\prime}$MnNiO$_6$ compounds (${A^\prime}$ = Cr, Mn, V, Co and Ni) are immense important for optoelectronic devices since they exhibit adequate semiconducting band gaps with highly positioned $d$-bands near the Fermi level. This can be facilitated for visible light absorption. Further, La${A^\prime}$MnNiO$_6$ DDPOs are expected to be good candidates for spintronic devices due to suppressed electron transfer for the semiconducting energy gaps. The Monte Carlo simulations and orbital-decomposed exchange parameters and hence exchange mechanisms elucidate the root cause of enhanced magnetic transition temperatures of these compounds. Thermodynamic study unravels their possibility of formation at high temperature and pressure while dynamic study dictates their kinetic stability. Our theoretical investigation determines these five La${A^\prime}$MnNiO$_6$ DDPOs to be critical integration in the ferro/ferrimagnetic insulators with remarkably high transition temperatures.
%%%%%%%%%%%%%%%%%%%%%%%%%%%%
\section*{Data Availability Statement}
The data that support the results of this study are available from the corresponding
 author upon reasonable request.
\begin{acknowledgments}
M.S. acknowledges INSPIRE division, Department of Science and Technology, New Delhi-110 016, Government of India for his fellowship [IF170335]. S.B. and R.G. acknowledge SRMIST for their fellowship. S.G. acknowledges the DST-SERB Core Research Grant(File No. CRG/2018/001728)for funding. The authors sincerely acknowledge SRMIST HPCC for providing computational resources. D.W. acknowledges financial support from the Science and Technology Development Fund from Macau SAR (Project Nos. 0062/2023/ITP2 and 0016/2025/RIA1) and the Macao Polytechnic University (Project No. RP/FCA-03/2023).
\end{acknowledgments}
%%%%%%%%%%%%%%%%%%%%%%%%%%%%%%%%%%%%%%%%%%%%%%%%%%%%%%%%%%%%%%%%%%%%%
\bibliographystyle{apsrev4-2}

%\bibliography{Monirul}
%%%%%%%%%%%%%%%%%%%%%%%%%%%%%%%%%%%%%%%%%%%%%%%%%%%%%%%%%%%%%%%%%%%%%
\end{document}